\documentclass[article]{IEEEtran}
\usepackage{cite}
\usepackage{booktabs}
\usepackage[utf8]{inputenc}
\usepackage{graphicx}
\usepackage{acronym}
\usepackage{authblk}
\usepackage{blindtext}
\usepackage{lettrine}
\usepackage [fleqn] {amsmath}
\usepackage{cases}
\usepackage{hyperref}
\usepackage{xcolor}
\ifCLASSOPTIONcompsoc
    \usepackage[caption=false, font=normalsize, labelfont=sf, textfont=sf]{subfig}
\else
\usepackage[caption=false, font=footnotesize]{subfig}
\fi
\usepackage{tabu}

\acrodef{DL}[DL]{Deep Learning}
\acrodef{CDF}[CDF]{Cumulative Distribution Function}
\acrodef{DNN}[DNN]{Deep Neural Network}
\acrodef{SoC}[SoC]{System on a Chip}
\acrodef{MDLS}[MDLS]{Memristive Deep Learning Systems}
\acrodef{VMM}[VMM]{Vector-Matrix Multiplication}
\acrodef{CNN}[CNN]{Convolutional Neural Network}
\acrodef{SC}[SC]{Stochastic Computing}
\acrodef{CBRAM}[CBRAM]{Conductive Bridging RAM}
\acrodef{SGD}[SGD]{Stochastic Gradient Descent}
\acrodef{ADC}[ADC]{Analog to Digital Converter}
\acrodef{LFSR}[LFSR]{Linear Feedback Shift Register}
\acrodef{CMOS}[CMOS]{Complementary Metal Oxide Semiconductor}
\acrodef{RRAM}[RRAM]{Resistive Random Access Memory}
\acrodef{MRAM}[MRAM]{Magnetoresistive Random Access Memory}
\acrodef{IoT}[IoT]{Internet of Things}
\acrodef{MAC}[MAC]{Multiply and Accumulate}
\acrodef{RNG}[RNG]{Random Number Generator}
\acrodef{APC}[APC]{Approximate Parallel Counter}
\acrodef{LDMOS}[LDMOS]{Laterally-Diffused Metal Oxide Semiconductor}

\begin{document}
\title{Memristive Stochastic Computing for Deep Learning Parameter Optimization}
\author{\IEEEauthorblockN{Corey Lammie\textsuperscript{\href{https://orcid.org/0000-0001-5564-1356}{\includegraphics[scale=0.035]{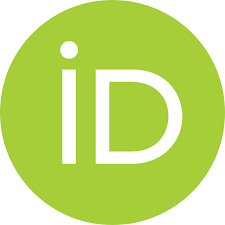}}},~\IEEEmembership{Student Member,~IEEE}, Jason K. Eshraghian\textsuperscript{\href{https://orcid.org/0000-0002-5832-4054}{\includegraphics[scale=0.035]{ORCID.png}}},~\IEEEmembership{Member,~IEEE}, Wei D. Lu\textsuperscript{\href{https://orcid.org/0000-0003-4731-1976}{\includegraphics[scale=0.035]{ORCID.png}}},~\IEEEmembership{Fellow,~IEEE}, and Mostafa Rahimi Azghadi\textsuperscript{\href{https://orcid.org/0000-0001-7975-3985}{\includegraphics[scale=0.035]{ORCID.png}}},~\IEEEmembership{Senior Member,~IEEE}}
\thanks{Corey Lammie and M.~Rahimi Azghadi are with the College of Science and Engineering, James Cook University, Australia (e-mail: \{\href{mailto:corey.lammie@jcu.edu.au}{corey.lammie}, \href{mailto:mostafa.rahimiazghadi@jcu.edu.au}{mostafa.rahimiazghadi}\}@jcu.edu.au).

Jason K. Eshraghian and Wei D. Lu are with the Department of Electrical Engineering and Computer Science, University of Michigan, USA (email: \{\href{mailto:jasonesh@umich.edu}{jasonesh}, \href{mailto:wluee@umich.edu}{wluee}\}@umich.edu).

\hspace{-1em}\rule{3cm}{0.5pt} \newline \textcopyright  \hspace{1pt} 2021 IEEE. Personal use of this material is permitted. Permission from IEEE must be obtained for all other uses, in any current or future media, including reprinting/republishing this material for advertising or promotional purposes, creating new collective works, for resale or redistribution to servers or lists, or reuse of any copyrighted component of this work in other works.
}}

\markboth{ACCEPTED BY IEEE TRANSACTIONS ON CIRCUITS AND SYSTEMS PART II: EXPRESS BRIEFS, 2021}%
	{Lammie, \MakeLowercase{\textit{et al.}}: \MyPaperTitlePlain}
	
\maketitle

\begin{abstract}
    \ac{SC} is a computing paradigm that allows for the low-cost and low-power computation of various arithmetic operations using stochastic bit streams and digital logic. In contrast to conventional representation schemes used within the binary domain, the sequence of bit streams in the stochastic domain is inconsequential, and computation is usually non-deterministic. In this brief, we exploit the stochasticity during switching of probabilistic \ac{CBRAM} devices to efficiently generate stochastic bit streams in order to perform \ac{DL} parameter optimization, reducing the size of \ac{MAC} units by 5 orders of magnitude. We demonstrate that in using a 40-nm \ac{CMOS} process our scalable architecture occupies 1.55mm$^2$ and consumes approximately 167$\mu$W when optimizing parameters of a \ac{CNN} while it is being trained for a character recognition task, observing no notable reduction in accuracy post-training.
\end{abstract}


\section{Introduction}
\lettrine{E}{mbedded} \ac{RRAM}-based neuromorphic and \ac{DL} accelerators have attracted significant attention due to their promise to revolutionize computing~\cite{RahimiAzghadi2020}. Such devices are capable of in-memory computation, and can be used to perform near-sensor high-speed and low-power computation at the \ac{IoT} edge~\cite{Krestinskaya2020}. 

Current research efforts are primarily focused towards the realization of scalable and reliable memristive architectures for in-memory computing applications~\cite{RahimiAzghadi2020,Mehonic2020}. For example, memristive crossbar architectures can be used to efficiently implement \ac{MAC} or dot-product accelerators to perform 2D \acp{VMM}, which are prominent in both neuromorphic and \ac{DL} systems, in $\mathcal{O}(1)$~\cite{Lammie2019}. Resistive memories, however, are still considered an emerging technology~\cite{Chen2020} that are prone to device-to-device variability, endurance challenges, and stochastic behavior~\cite{Li2018}, which make reaching the maximum gain of \ac{RRAM} technology, currently unfeasible.

While the commercial long-term viability of using \ac{RRAM} devices for such purposes is yet to be properly determined~\cite{Mittal2019}, rather than treating the non-ideal stochastic behavior of \ac{RRAM} devices as a hindrance, some researchers have sought to exploit the well-characterized stochasticity of \ac{RRAM} devices for alternative applications such as chaos~\cite{Muthuswamy2009,Zheng2019} and random number generation~\cite{Jiang2017,Zhang2017}. Traditionally, large hardware costs associated with stochastic number generation have hindered stochastic processors, as they require large bit stream lengths to mitigate undesirable computational errors~\cite{deAguiar2015}, rendering them largely ineffective for most applications. In~\cite{Knag2014}, a hybrid \ac{CMOS}-memristor stochastic processor was proposed, which used digital \ac{CBRAM} devices for efficient stochastic number generation. It was demonstrated that when using large bit-stream lengths, the processor was able to perform gradient descent optimization and k-means clustering in a low-power and high-speed mode of operation.

\begin{figure}[!t]
    \centering
    \includegraphics[width=0.5\textwidth]{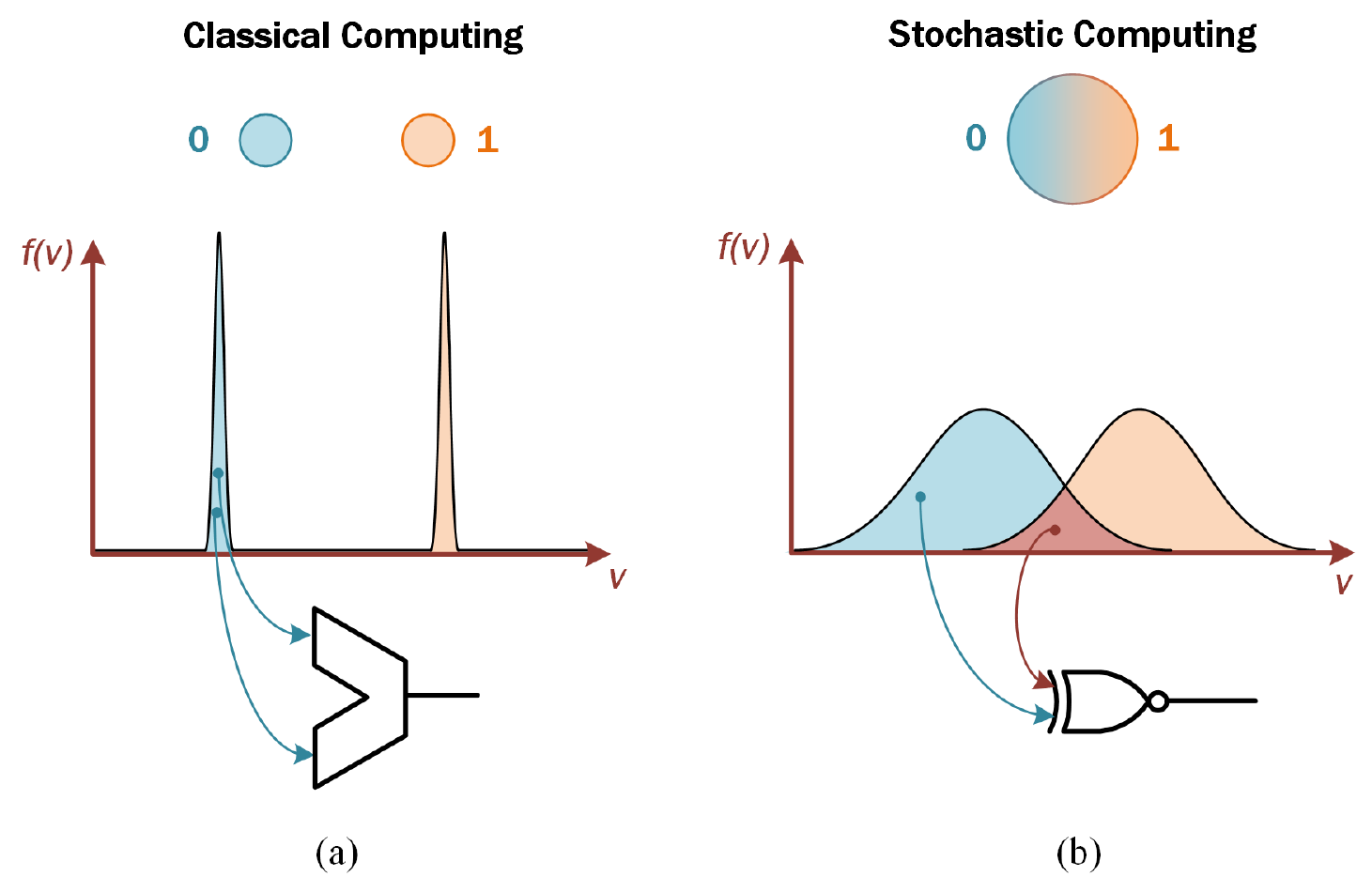}
    \caption{Conceptual difference between classical and stochastic computing in terms of probability density functions. (a) Classical or deterministic computing requires well-defined margins. 
   (b) Stochastic computing is error-tolerant, and can thus exploit quantum models of uncertainty that occur at the particle level. This simplifies the hardware required for downstream processing. In our case, a large arithmetic logic unit containing a multiplier can be replaced by a single logic gate.}
    \label{fig1}
\end{figure}

In this brief, we expand upon efforts in~\cite{Knag2014}, and exploit the well characterized switching stochasticity of probabilistic \ac{CBRAM} devices to efficiently generate stochastic bit streams in order to perform deep learning parameter optimization using a hybrid \ac{CMOS}-memristor stochastic processor. By nature, such a processor is highly tolerant to external noise and relaxes many of the stringent hardware requirements needed in generating distinct voltage levels (Fig.~\ref{fig1}). While most prior \ac{DL}-based \ac{SC} research applies \ac{SC} to the feed-forward processing (inference) stage, as it is known to be tolerant to noise, we instead explore if it is at all possible to perform parameter optimization, that typically requires high precision, using probabilistic bits. Our specific contributions are as follows:

\begin{enumerate}
    \item We are the first to exploit the switching stochasticity of \ac{CBRAM} devices to perform deep learning parameter optimization using \ac{SC};
    \item We evaluate our architecture by training a \ac{DNN} for the MNIST character recognition task using Mini-Batch \ac{SGD} and Mini-Batch \ac{SGD} with Momentum;
    \item We investigate the tradeoff between latency, area and power consumption, and demonstrate that our architecture can be used to reduce the size of MAC units.
\end{enumerate}

\section{Preliminaries}
\subsection{Stochastic Computing}

\begin{figure}[!b]
    \centering
    \includegraphics[width=0.45\textwidth]{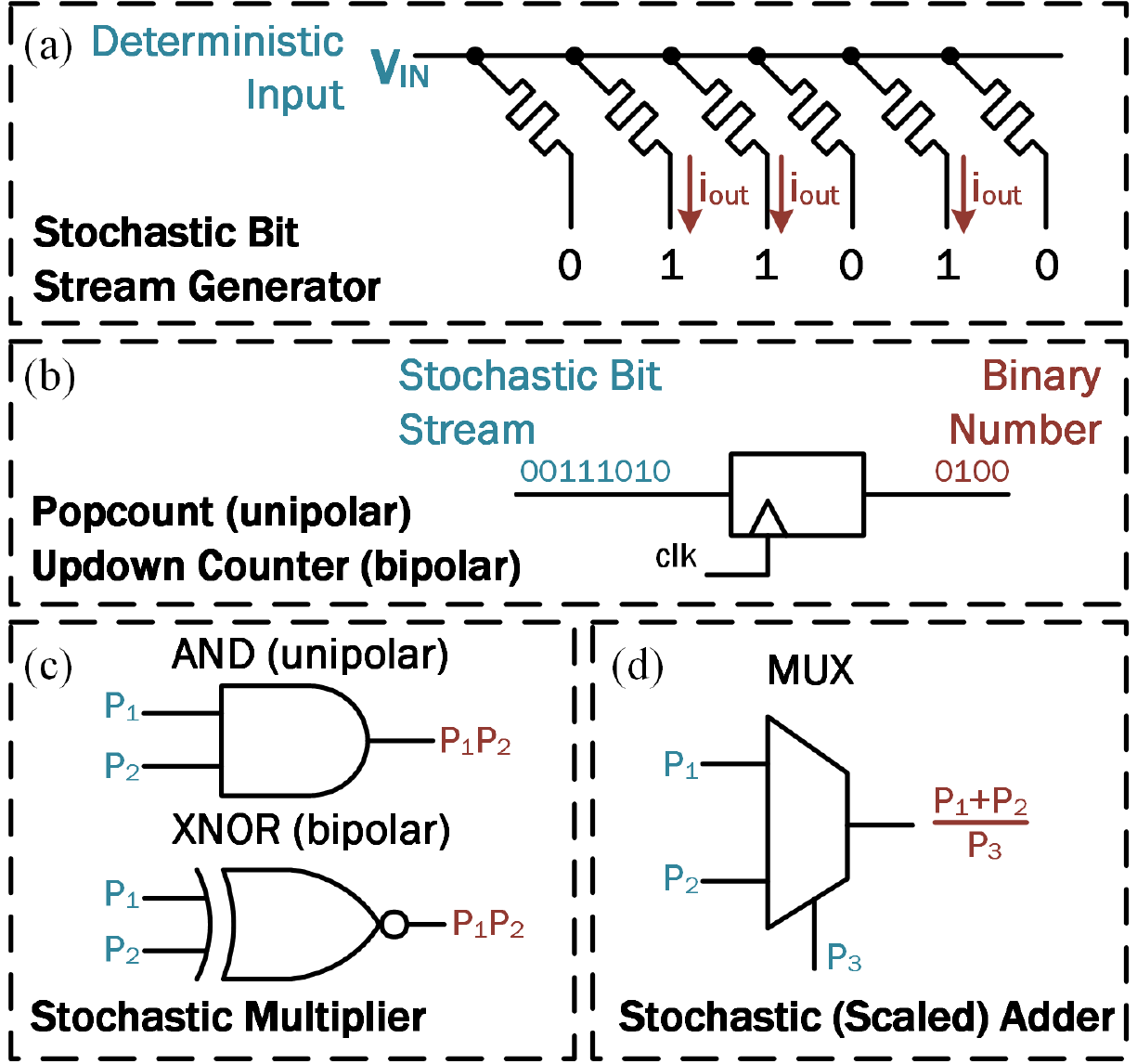}
    \caption{Stochastic computing blocks for (a) bit stream generation, (b) stochastic bit stream to binary converter, (c) multiplication, and (d) scaled addition.}
    \label{fig:stochastic_computing_blocks}
\end{figure}

In \ac{SC}, numbers are represented using bit streams. The frequency of 0s and 1s in stochastic bit streams determines their value depending on a range, $R$, denoted as the priori. The priori can be unipolar, i.e., $R \in [0,1]$ to represent unsigned operands, or bipolar to represent signed operands, i.e., $R \in [-1, 1]$. The value represented by unipolar bit streams can be determined using the mean, $x$, whereas the value represented by bipolar bit streams can be determined using $2x - 1$. Conventional \ac{SC} circuits use \acp{LFSR} to generate pseudo-random numbers for stochastic bit stream generation, and popcount and up-down counter circuits to decode bit streams. \ac{SC} blocks for bit stream generation, addition, multiplication, and stochastic bit stream to binary converters are depicted in Fig.~\ref{fig:stochastic_computing_blocks}.

\begin{figure}[!b]
    \centering
    \includegraphics[width=0.45\textwidth]{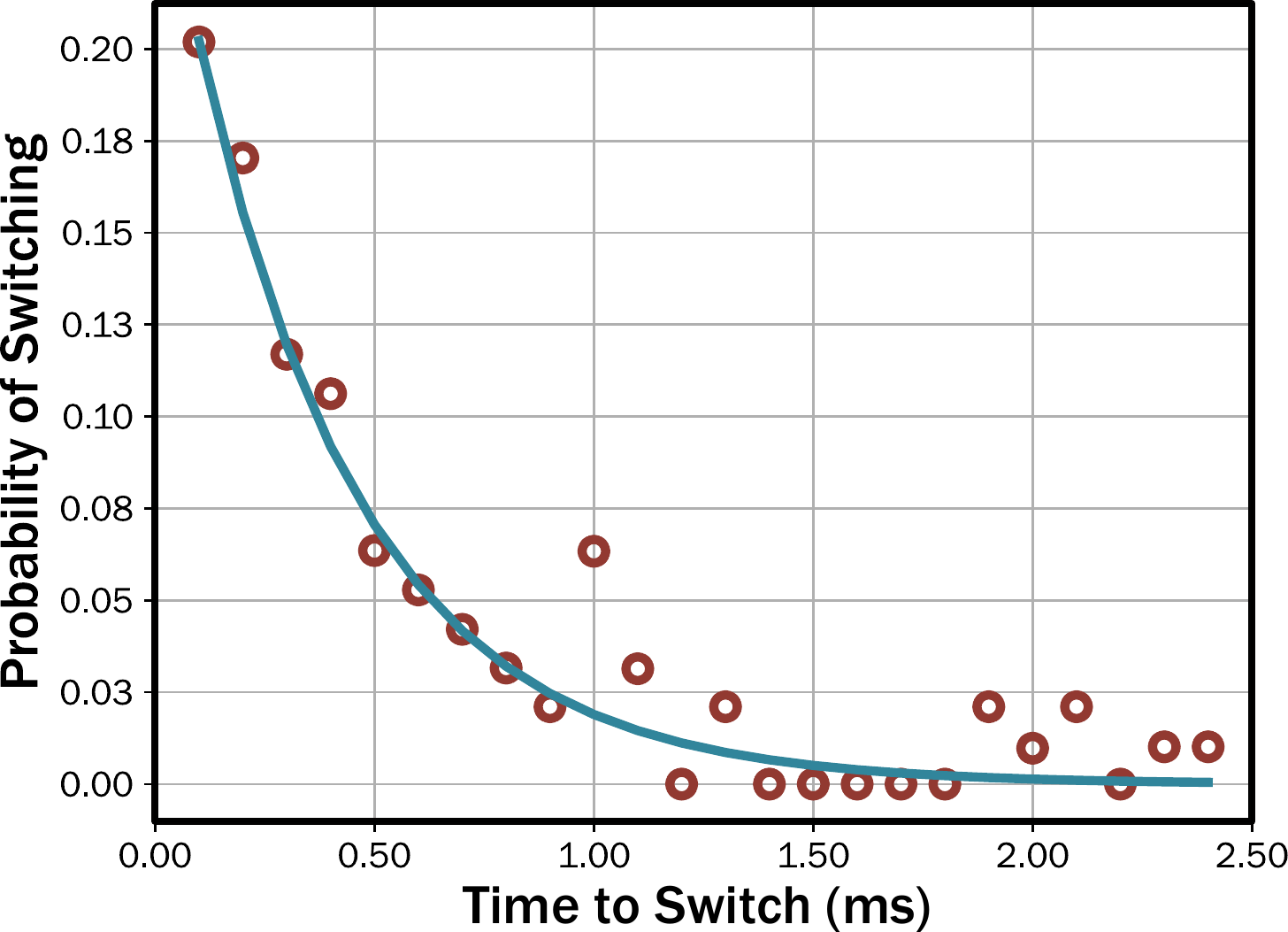}
    \caption{Distribution of switching time of a fabricated \ac{CBRAM} device~\cite{Gaba2013} under an applied voltage of 4.5 V. Switching events (red circles) fit a Poisson distribution (blue line), $P(t) = (\Delta t/\tau) e^{-t/\tau}$ for $\Delta t$ = 0.5, $V = 0.4$ and $\tau = 0.38$ms.}
    \label{fig:device_characteristics}
\end{figure}

\begin{figure*}[t]
    \centering
    \includegraphics[scale=0.75]{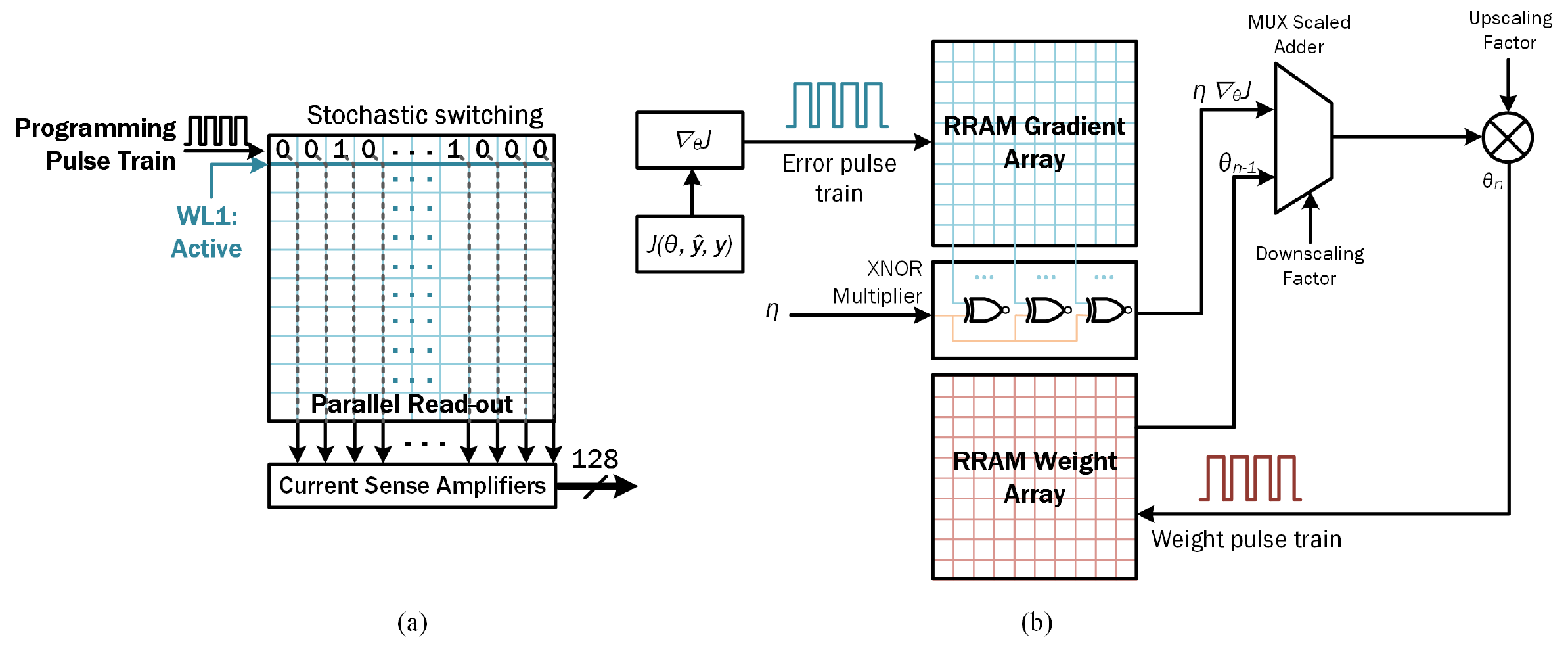}
    \caption{Native stochastic parameter update process. (a) The word line (WL) is activated to open one row at a time. A pulse train is applied to induce stochastic switching in the RRAM. The current sense amplifiers are used to sense which devices are drawing current, indicating which cells have been switched on.  (b) The parallel-generated bit stream is XNOR multiplied by the negative learning rate and added to the previous weights to perform a single parameter update. The overhead of requiring an additional array to compute the weight bit stream is partially offset by removing the need for \acp{ADC}.}
    \label{fig:architecture}
\end{figure*}

\subsection{Memristor-based Native Stochastic Computing}
Digital memristor devices have a large dynamic range between logic levels that is associated with the formation and rupture of a dominant, nanoscale conducting filament~\cite{Gaba2013}. On account of device-device variation, experimental studies have demonstrated that the switching time between logic levels is stochastic, and that the time-to-switch can be accurately modelled using a Poisson distribution~\cite{Sun2019,Naous2016}, as depicted in Fig.~\ref{fig:device_characteristics}. By exploiting this property, both unipolar and bipolar stochastic bit streams can be efficiently generated using digital memristors by applying programming pulses with variable pulse widths to memristor cells. Memristor-based native \ac{SC} systems can be realized by using memristor cells for stochastic bit stream generation, and using \ac{CMOS} for stochastic arithmetic circuits.

\subsection{Deep Learning Parameter Optimization}
\ac{SGD}~\cite{Ruder2016} is a parameter optimization method that is described by (\ref{eq:sgd}), which is commonly used to train \acp{DNN}. 
\begin{equation}\label{eq:sgd}
    \theta_n = \theta_{n-1} - \eta\nabla_\theta J(\theta, y'_i, y_i), 
\end{equation}
where, \noindent $\eta$ is the learning rate, $\theta$ denotes a trainable parameter, $y_i$ represents the class label, $y'_i$ denotes the predicted class label, and $J(\theta, y'_i, y_i)$ describes the objective function. Momentum~\cite{Sutskever2013}, or SGD with momentum, described by (\ref{eq:momentum}), is a method that improves \ac{SGD} and accelerates gradients in relevant directions towards a local minima by dampening oscillations using $\gamma$ and $v$, i.e., momentum and velocity parameters.
\begin{equation}\label{eq:momentum}
    \begin{aligned}
        v_n &= \gamma v_{n-1} + \eta\nabla_\theta J(\theta, y'_i, y_i), \\
        \theta_n &= \theta_{n-1} - v_n
    \end{aligned}
\end{equation}

Mini-Batch \ac{SGD} and Mini-Batch Momentum are variations of the \ac{SGD} and Momentum optimization algorithms which split training datasets into small batches that are used to perform parameter optimization. From this stage forward, we implicitly refer to mini-batch variations.

\begin{table}[!b]
\centering
\caption{Adopted Network Architecture\protect\footnotemark}
\label{table:network_architecture}
\begin{tabu} to 0.5\textwidth {p{0.3\textwidth}X[r]}
\toprule
 \textbf{Layer}  & \textbf{Output Shape}  \\ 
\midrule
 Convolutional, $f = 10, k=5, s=1, p=0$  & $(10 \times 24 \times 24)$  \\
Max Pooling, $k=2, p=2$ & $(10 \times 12 \times 12)$  \\
Convolutional, $f = 20, k=5, s=1, p=0$~~ & $(20 \times 8 \times 8)$  \\
Max Pooling, $k=2, p=2$ & $(20 \times 4 \times 4)$  \\
Fully Connected, $N = 50$ & $(50)$  \\
Fully Connected, $N = 10$ & $(10)$  \\
\bottomrule
\end{tabu}
\end{table}

\section{Proposed Architecture}
A block diagram of our proposed architecture is depicted in Fig.~\ref{fig:architecture}. We confine operating in the stochastic domain to the parameter update stage during training. The architecture consists of scalable crossbar tiles of probabilistic \ac{CBRAM} devices (\ac{RRAM} input and weight arrays) and \ac{CMOS} systolic stochastic arithmetic circuits (multipliers and scaled adders). Stochastic bit streams are generated by writing pulse trains to columns of 2D crossbar architectures. Given an applied voltage, $V$, and a programming pulse width, $t$, the switching probability can be expressed using (\ref{eq:prob})
\begin{equation}\label{eq:prob}
    P(t, V) = 1 - e^{(-te^{V/V_0}) / \tau_0},
\end{equation}
\noindent where $V_0$ and $\tau_0$ are fitting parameters~\cite{Knag2014}. By fixing the applied voltage and varying the programming pulse width applied to each column, bipolar bit streams can be efficiently generated in-memory. The bits are read out from the array in parallel using current sense amplifiers (Fig.~\ref{fig:architecture}(a)). 

\section{Training and Validation Methodologies}
To evaluate our architecture, we trained a \ac{CNN} described in Table~\ref{table:network_architecture} using MNIST. All convolutional, and the first fully connected layer were sequenced with batch normalization layers, and the ReLU activation function was used for all layers. All networks were trained for 10 epochs with a batch size of 256 and a fixed learning rate. Gradients were clipped between -1.0 and 1.0. Cross entropy loss was used in conjunction with \ac{SGD} and \ac{SGD} with momentum. For all networks trained using momentum, Nesterov was disabled, and a fixed momentum value of 0.9 was used, which has demonstrated significant performance for a variety of \ac{DL} tasks~\cite{Sutskever2013}.

\footnotetext{For each convolutional and pooling layer, $f$ denotes the number of filters, $k$ determines the filter size, $s$ is the stride length, and $p$ denotes the padding. $N$ is the number of output neurons for each fully connected layer.}

\section{Results}
\begin{figure*}[!t]
    \centering
    \includegraphics[width=1\textwidth]{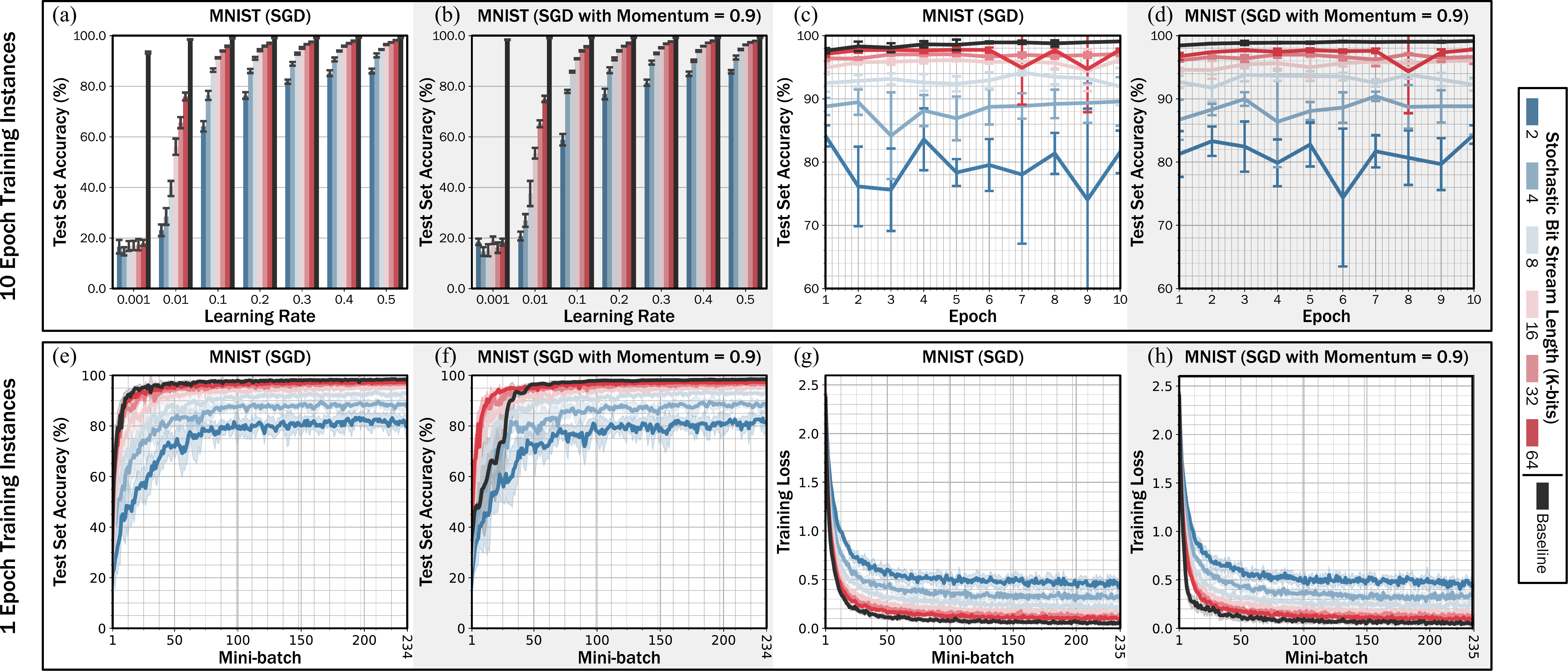}
    \caption{The MNIST test set accuracy across seperate training instances of 10 epochs (a--d) and 234 mini-batches (1 epoch) (e--h), respectively, for \acp{CNN} trained using our native SC-based parameter optimization with different learning rates for (a) SGD, and (b) SGD with momentum; for a fixed learning rate with (c) SGD, and (d) SGD with momentum; for a fixed learning rate with (e) SGD, and (f) SGD with momentum. The training loss during the first training epoch for (g) SGD, and (h) SGD with momentum. In (c--h) a learning rate of 0.5 was used. Each baseline implementation uses 32-bit floating point weights.}
    \label{fig:results}
\end{figure*}

\subsection{Performance}
The MNIST test set accuracy for all training epochs, and the MNIST test set accuracy and training loss for all mini-batches during the first training epoch are reported in Fig.~\ref{fig:results}, for network architectures trained using our native SC-based parameter optimization with \ac{SGD} and \ac{SGD} with momentum. These are also shown for a baseline implementation using conventional training. The learning rate was varied from 0.01 -- 0.5 and the stochastic bit stream length (N$_{\rm bit}$) was varied from 2-Kbits -- 64-Kbits, near the range investigated in~\cite{Knag2014}. Each training instance was repeated 10 times, and the mean and standard deviation across instances were determined. From Fig.~\ref{fig:results}, it can be observed that for both \ac{SGD} and \ac{SGD} with momentum a bit stream length larger than 8-Kbit and a learning rate of $\geq$0.1 is required for stable training performance. We attribute the performance degradation when smaller learning rates are used to the fact that positive values near zero can be encoded negatively in the stochastic domain, and note that such implementations may eventually converge if trained for $\gg$10 epochs.

\subsection{Power and Area Requirements}
Power and area estimates were calculated based on a 40~nm CMOS process integrated with RRAM in the back end of the line \cite{Wang2019}. Specifically, a pre-configured IP is used for the encoder and decoder, and a full-custom approach was taken to laying out the RRAM array, sense-amplifiers, registers, MUX-based scaled adder, and XNOR accumulator, with verified DRC/LVS compliance. Power draw is modified to suit the device distribution in Fig.~\ref{fig:device_characteristics}. The area of each 128$\times$128 RRAM tile is 2.77$\times$10$^3\mu$m$^2$. Assuming that the entire bit stream must be accessible at once, and that only one row may be read out at a time, a bit stream length of 16-Kbits requires 2$^7$ tiles. The gradient bit stream is multiplied by a learning rate bit stream of equivalent length and subtracted from the weight. To improve throughput, the number of arrays is doubled to allocate half of the tiles to be reset while the other half are generating bit streams. This gives a total of 2$^{9}$=512 RRAM tiles, occupying a total area of 1.42mm$^2$.

Static power consumption is estimated based on the presumption that half of the tiles are generating bit-streams while the other half are being reset, one row at a time. Balancing the latency between reading and resetting is feasible because reading requires stochastic programming. 
As a conservative estimate, we assume an input voltage of 4.5V, which is tolerable at 40nm using \ac{LDMOS} transistors, that draw approximately 10nA from a device that is on, and 100pA from a device that is off \cite{Gaba2014}. The corresponding static power dissipation is 45nW and 0.45nW, respectively. The power consumed for one read out can be measured by (\ref{eq:power}):
\begin{equation}\label{eq:power}
    P_{\rm read} = N_{\rm bit}[P_{\rm on}\mathbf{E}(\mathbf{\nabla_\theta J}) + P_{\rm off} (1-\mathbf{E}(\mathbf{\nabla_\theta J}))],
\end{equation}

\noindent where N$_{\rm bit}$ is the bit stream length, P$_{\rm on}$ and P$_{\rm off}$ are the power dissipated from an on and off device, and $\mathbf{E}(\mathbf{\nabla_\theta J})$ is the expected value of the gradient $\nabla_\theta J(\theta, y'_i, y_i)$. We measured $\mathbf{E}(\mathbf{\nabla_\theta J})$=0.5367 at the start of the training process. Gradient updates are generally larger at the start of training than at convergence. Therefore, more cells are switched on for steeper gradients. This means that worst-case power consumption should be measured at the start of training. For normalized bipolar weights, $\mathbf{E}(\mathbf{\theta})$=0.5. Power dissipation from the RRAM array is estimated to be 43.0~$\mu$W per gradient, and 40.6~$\mu$W per weight. These estimates are doubled to account for additional tiles used for resetting and enhancing throughput, giving a total result of 167~$\mu$W.

The layout of the peripheral CMOS circuitry was used to generate accurate area and power estimates. The area of a single XNOR gate is 670nm$\times$355nm. The pitch of the RRAM array is approximately 410nm. The XNOR multiplier can be oriented such that it is pitch-matched to the RRAM columns. It can then fit under the bottom row without additional area overhead. The total area occupation of the XNOR gate of 0.031mm$^2$ can thus be merged with the RRAM area. The total area overhead from accumulation consists of the MUX-based scaled adder and a register which consume 0.0735mm$^2$. Each sense-amplifier built from a pair of cross-coupled inverters occupy 0.41$\mu$m$^2$. Although the shorter dimension can be optimized to be pitch-matched to the array, a reference signal is required to distinguish on and off cells. This reference must be obtained from an adjacent column, which requires time multiplexing the bit stream generation process into two steps. With pairwise column sharing, the total area of all current sense amplifiers is 0.0267mm$^2$. Time multiplexing also halves the maximum static power dissipation to 83.5~$\mu$W. Total static power dissipation of the CMOS elements is from subthreshold current draw, and thus dominated by resistive dissipation. 

\section{Discussion}
The length of the parallel-generated bit stream may cause large power draw as throughput is increased. But this drawback is offset by three factors. Firstly, the \ac{MAC} process of non-quantized weights can now be completed in one single step using an XNOR gate and a multiplexer (with a simulated post-layout delay of 58~ps). A 16-bit \ac{MAC} unit designed in a comparable process \cite{Shoba2017} is approximately five orders of magnitude larger and two orders slower than an XNOR gate and a MUX. Secondly, the power estimates are derived from the peak value at the end of bit stream generation. All devices are initialized to be off and only switch on during the course of the time to switch (Fig.~\ref{fig:device_characteristics}). This peak only occurs at the end of the read out process. Finally, sampling inputs and weights from a normal distribution tracks gradient descent pathways that avoid saddle points that riddle high-dimensional problems. 

Numerous options are available to further optimize the power dissipation. Significant resources have been dedicated to developing low-voltage memristors well below the supply limitation of the 40nm process used. At one extreme end, the work in~\cite{Zhu2020} uses 100~mV programming pulses. This would reduce resistive dissipation by a larger factor, though must be balanced with the average switching time. Our work has shown that lengthy bit streams are required to obtain acceptable convergence in the training process. This is on par with literature \cite{Knag2014}, however, we expect that the use of quantization-aware training can effectively place a constraint upon the set of permissible values, thus reducing the bit length.

Alternatively, the generation of long bit streams can be time multiplexed to free up arrays to increase throughput. Although this slows down bit stream generation, we have eliminated the large delays associated with \ac{MAC} operations by one order of magnitude \cite{Shoba2017}. \ac{SC} also uses noise to avoid being trapped by saturated gradients during training, known to significantly slow down the training process in conventional computing. Endurance concerns can be overcome by substituting \ac{RRAM} for \ac{MRAM}, which exhibits endurance of over $10^{15}$ cycles, compared to that of \ac{RRAM} ($\approx 10^5-10^{12}$ cycles) with faster write times, at the cost of increased fabrication complexity.

\section{Conclusion}
In this brief, we demonstrated that it is indeed possible to perform \ac{DL} parameter optimization using stochastic bits by exploiting the stochasticity during switching of probabilistic \ac{CBRAM} devices to efficiently generate stochastic bit streams. This new insight to stochastic computing is valuable for the following reasons:

\begin{enumerate}
    \item For an end-to-end stochastic computing system, the gradient update step can share resources with the feed-forward step; the alternative to long bit-streams would be a floating point unit, which offsets this disadvantage;
    \item The multiply-and-accumulate steps now rely only on two combinational logic gates. This means the propagation delay for MAC is reduced by orders of magnitude.
\end{enumerate}
An investigation of switching probability variation on account of device-to-device variation and an end-to-end timing analysis at the circuit and system level for a variety of configurations to train more complex networks using larger data sets and forms the basis of future work. 

\bibliographystyle{IEEEtran}

\end{document}